\documentclass[twocolumn,prb,amsmath,amssymb,floatfix,superscriptaddress,showpacs]{revtex4}
\usepackage{graphicx}
\usepackage{dcolumn}
\usepackage{bm}
\usepackage{times}
\begin{document}


\title{A two-component model of the neutron diffuse scattering in the relaxor ferroelectric PZN-4.5\%PT}

\author{Zhijun Xu}
\affiliation{Condensed Matter Physics and Materials Science
Department, Brookhaven National Laboratory, Upton, New York 11973,
USA} \affiliation{Department of Physics, City College of New York,
New York, New York 10033, USA}
\author{Jinsheng Wen}
\affiliation{Condensed Matter Physics and Materials Science Department, Brookhaven National Laboratory, Upton, New York 11973,
USA}\affiliation{Department of Materials Science and Engineering, Stony Brook University, Stony Brook, New York 11794, USA}
\author{Guangyong Xu}
\affiliation{Condensed Matter Physics and Materials Science Department, Brookhaven National Laboratory, Upton, New York 11973,
USA}
\author{C. Stock}
\affiliation{ISIS Facility, Rutherford Appleton Laboratory, Didcot, OX11 0QX, UK}
\author{J. S. Gardner}
\affiliation{Indiana University, 2401 Milo B. Sampson Lane,
Bloomington, Indiana 47408, USA} \affiliation{NIST Center for
Neutron Research, National Institute of Standards and Technology,
Gaithersburg, Maryland 20899, USA}
\author{P. M. Gehring}
\affiliation{NIST Center for Neutron Research, National Institute of Standards and Technology, Gaithersburg, Maryland 20899, USA}

\date{\today}

\begin{abstract}
We report measurements of the neutron diffuse scattering in a single
crystal of the relaxor ferroelectric material
95.5\%Pb(Zn$_{1/3}$Nb$_{2/3}$)O$_3$-4.5\%PbTiO$_3$ (PZN-4.5\%PT).
Our results suggest that the nanometer scale structure in this
compound exhibits both $\langle100\rangle$ and $\langle110\rangle$
polarizations, which contribute to different portions of the total
diffuse scattering intensity.  These contributions can be
distinguished by the differing responses to an electric field
applied along [001].  While diffuse scattering intensities
associated with $\langle110\rangle$ (T2-type) polarizations show
little to no change in a [001] field, those associated with
$\langle100\rangle$ (T1-type) polarizations are partially suppressed
by the field at temperatures below the Curie temperature
\emph{T$_C$} $\sim 475$\,K.  Neutron spin-echo measurements show
that the diffuse scattering at (0.05,0,1) is largely dynamic at high
temperature and gradually freezes on cooling, becoming mostly static
at 200\,K.

\end{abstract}

\pacs{77.80.-e, 77.84.Dy, 61.12.Ex}

\maketitle

\section{Introduction}

The study of relaxors has grown dramatically over the past two
decades.  While many interesting properties distinguish relaxors
from conventional ferroelectrics, the hallmark of relaxors is a
highly frequency-dependent dielectric response that peaks broadly at
a temperature that is unrelated to any structural phase
transition.~\cite{park1997,uchino1996,service1997}  Although the
origin of many relaxor properties are still not well understood,
most researchers tend to agree that the chemical short-range order
in these materials, which are primarily compositionally disordered
oxides, plays a key role in determining the bulk
response.~\cite{cross1987,Burton2006}  In the case of the well-known
lead-based, perovskite ABO$_3$ relaxors
Pb(Mg$_{1/3}$Nb$_{2/3}$)O$_3$~(PMN) and
Pb(Zn$_{1/3}$Nb$_{2/3}$)O$_3$~(PZN),~\cite{park1997,Kuwata1981} the
mixture of 2$^+$ and 5$^+$ B-site cations leads to local charge
imbalances that create random fields that destroy long-range polar
order. On the other hand short-range polar order, more commonly
known as polar nanoregions (PNR), appears at temperatures well above
the temperature at which the dielectric susceptibility reaches a
maximum ($T_{max}$).~\cite{burns1983} There have been substantial
experimental evidences exist that suggest the PNR influence various
bulk properties of relaxors such as the thermal expansion in
PMN,~\cite{Gehring2009} the piezoelectric coefficients in PMN doped
with PbTiO$_3$ (PMN-$x$PT),~\cite{Matsuura2006} and the transverse
acoustic phonon lifetimes in PZN doped with 4.5\%\ PbTiO$_3$
(PZN-4.5\%PT).~\cite{gxu2008nm}  However, unlike the chemical
short-range order, which is quenched and thus does not change with
temperature, the polar short-range order (PNR) is sensitive to both
temperature and external electric fields.  It has therefore been the
focus of numerous studies.

Different techniques including dielectric
spectroscopy,~\cite{cross1987,colla1998} Raman
scattering,~\cite{Welsch2009,Ohwa1998,Ohwa1999,Siny2000} and
piezoelectric force microscopy~\cite{Lehnen2001,Shvartsman2004} have
been used to explore the behavior of PNR in relaxor systems. Yet the
most effective and direct probe of PNR within bulk relaxor samples
is arguably obtained through measurements of the corresponding
diffuse scattering, which reflects the presence of short-range
ordered, atomic displacements.~\cite{gxu2010jpsj}  Many
neutron~\cite{gxu2004prb,Vakhrushev1995,hirota2002,Hiraka2004prb,Hlinka2003jpcm,Gvasaliya2004,La-Orauttapong2003,Dkhil2001,Jeong2005}
and x-ray diffuse~\cite{You1997,Takesue2001,gxu2006nm} scattering
studies have been performed to determine the structure,
polarization, and other properties of PNR.  It has been shown that
the diffuse scattering in pure PMN, PZN,  and their solid solutions
with low PbTiO$_3$ (PT) doping, exhibits very similar
behavior.~\cite{gxu20043d,Matsuura2006}  In all cases, the diffuse
scattering intensities are strongly anisotropic, varying with
Brillouin zone, and can be affected by an external electric field.
Most neutron and x-ray diffuse scattering studies also agree that
the diffuse scattering in both PMN-$x$PT and PZN-$x$PT systems
extends preferentially along $\langle110\rangle$ in reciprocal
space.  This diffuse scattering is rod-shaped but adopts an ``X'' or
butterfly shape when measured in the (HK0) scattering plane near
reflections of the form $(h00)$.  Detailed analysis indicates that
the butterfly-shaped diffuse scattering is associated with
short-range ordered, ionic displacements oriented along
$\langle1\overline{1}0\rangle$ ~\cite{gxu20043d} and couples
strongly to transverse acoustic (TA) phonons propagating along
$\langle110\rangle$ (TA$_2$ modes).~\cite{gxu2008nm,Stock2005}  For
this reason we shall refer to it as ``T2-diffuse scattering'' in
this article.  The T2-diffuse scattering intensities come from PNR
composed of ionic displacements that are neither purely strain (no
relative changes between A, B, and O site positions) nor purely
polar (only relative changes between A, B, and O site positions),
which suggests that the local atomic structure both within and
around the PNR is very complicated.~\cite{gxu2006prb}

\begin{figure}[ht]
\includegraphics[width=0.9\linewidth]{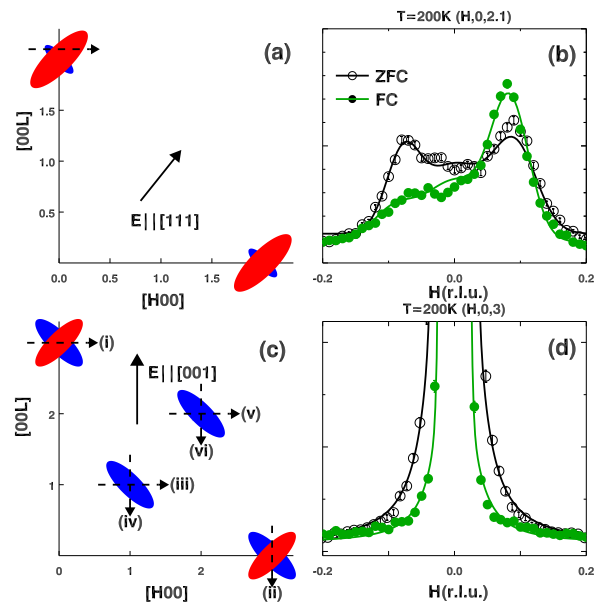}
\caption{(Color online) (a) Schematic diagram of the neutron
scattering measurements made in the (H0L) scattering plane with an
electric field applied along [111]. The large (red) and small (blue)
ellipses reflect how the T2-diffuse scattering intensity is
redistributed after the sample is field cooled (FC). (b)
Measurements of the diffuse scattering intensity near (002) along
the dashed arrow shown in panel (a) at 200\,K (from
Ref.~\onlinecite{gxu2008nm}).  Open circles represent zero-field
cooled (ZFC) data and closed circles represent FC data . (c) Same
schematic diagram as in panel (a) except with the electric field
applied along [001].  Dashed lines denote linear $q$-scans performed
across Bragg peak positions. (d) Linear $q$-scans at 200\,K along
[100] measured across (003) [see scan (i) in panel (c)].  Error bars
in (b) and (d) represent the square root of the number of counts.}
\label{fig:1}
\end{figure}

When an external electric field is applied along [111], the
T2-diffuse scattering intensities are redistributed between
different rods of diffuse scattering, while the overall diffuse
scattering intensity appears to be conserved [see panels (a) and (b)
of Fig.~\ref{fig:1}].~\cite{gxu2006nm}  However when an external
field is applied along [001], the T2-diffuse scattering is
essentially unchanged.~\cite{Wen2008apl}  At the same time, there is
evidence that suggests the presence of another type of diffuse
scattering different from the T2-diffuse scattering. For example,
Gehring {\it et al}.\ found that the diffuse scattering measured in
PZN-8\%PT near (003) along [100], i.\ e.\ in a transverse direction,
is strongly suppressed by an external [001] field, whereas that
measured near (300) along [001] remains
unaffected.~\cite{Gehring2004prb} This effect has been confirmed in
our measurements on PZN-4.5\%PT  [see panels (c) and (d) of
Fig.~\ref{fig:1}] and cannot be explained by a simple redistribution
of the T2-diffuse scattering.

In this article we report diffuse scattering measurements made under
a [001]-oriented electric field on the relaxor ferroelectric
PZN-4.5\%PT at reduced wavevectors $q$ offset from various Bragg
peaks along $\langle001\rangle$ in the (H0L) zone [refer to the
dashed lines in panel (c) of Fig.~\ref{fig:1}].  We show that, in
addition to the T2-diffuse scattering, which dominates the total
diffuse scattering intensity in most cases, there is another
component of diffuse scattering which is primarily distributed along
$\langle001\rangle$.  We shall refer to this component as
``T1-diffuse scattering'' because of its similarities in both
polarization and propagation direction to that of T1 phonon modes in
perovskite systems.  The application of a [001] electric field has
no obvious effect on the T1-diffuse scattering which is associated
with short-range ordered, ionic displacements polarized along [100],
but strongly suppresses the diffuse scattering which is associated
with short-range ordered, ionic displacements polarized along [001].
This effect is most prominent at temperatures slightly below $T_C
\sim 475$\,K.  In addition, spin-echo measurements performed on the
same sample show that the diffuse scattering has a large dynamic
component at high temperature (550\,K), which gradually freezes and
becomes almost fully static at 200\,K.

\begin{figure}[ht]
\includegraphics[width=\linewidth]{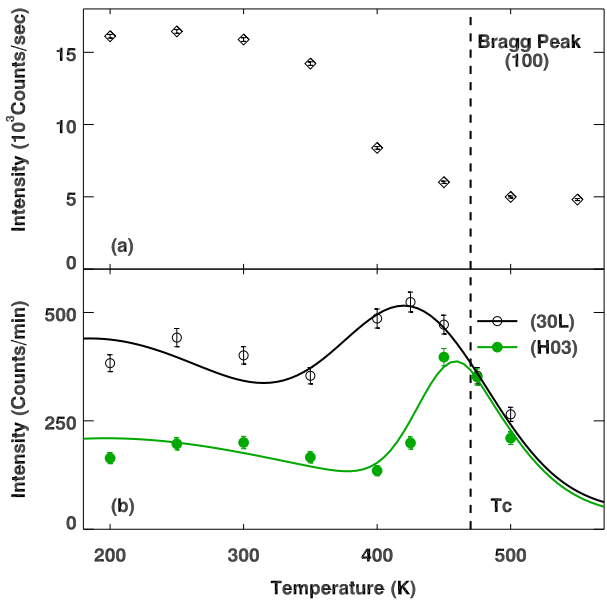}
\caption{(Color online) (a) (100) Bragg peak intensity versus
temperature. The intensity increase near $T_C$ is the result of a
release of extinction at the phase transition.  (b) Diffuse
scattering intensity measured at (3,0,0.06) (open black symbols) and
(-0.06,0,3) (closed green symbols) after cooling (FC) in a field
$E=4$\,kV/cm applied along [001].  The lines are guides to the eyes,
and the error bars represent the square root of the number of
counts.} \label{fig:2.5}
\end{figure}

\begin{figure}[ht]
\includegraphics[width=\linewidth]{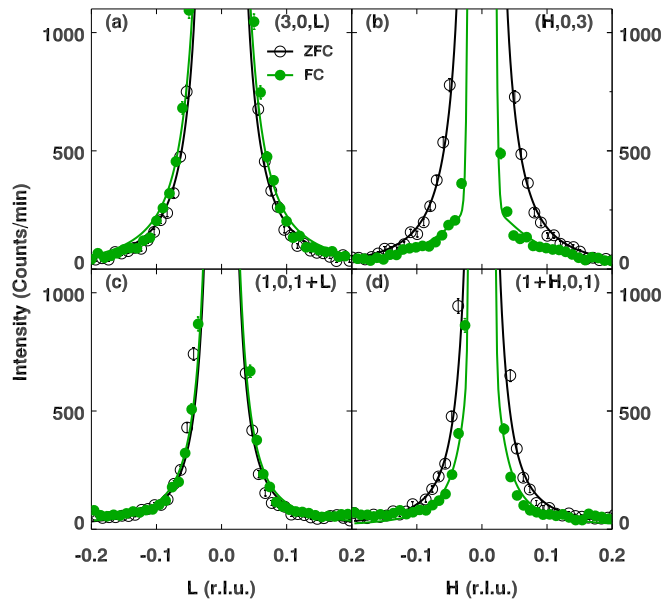}
\caption{(Color online) Diffuse scattering intensity measured at
400\,K.  ZFC data are shown as (black) open circles; FC data are
shown as (green) closed circles.  Data were taken around (a) (3,0,L)
[scan (ii) in panel (c) of Fig.~\ref{fig:1}]; (b) (H,0,3) [scan (i)
in panel (c) of Fig.~\ref{fig:1}]; (c) (1,0,1+L) [scan (iv) in panel
(c) of Fig.~\ref{fig:1}]; and (d) (1+H,0,1) [scan (iii) in panel (c)
of Fig.~\ref{fig:1}]. The solid lines are based on least square fits
to the data described in the text.  The error bars represent the
square root of the number of counts.}
\label{fig:2}
\end{figure}

\begin{figure}[ht]
\includegraphics[width=\linewidth]{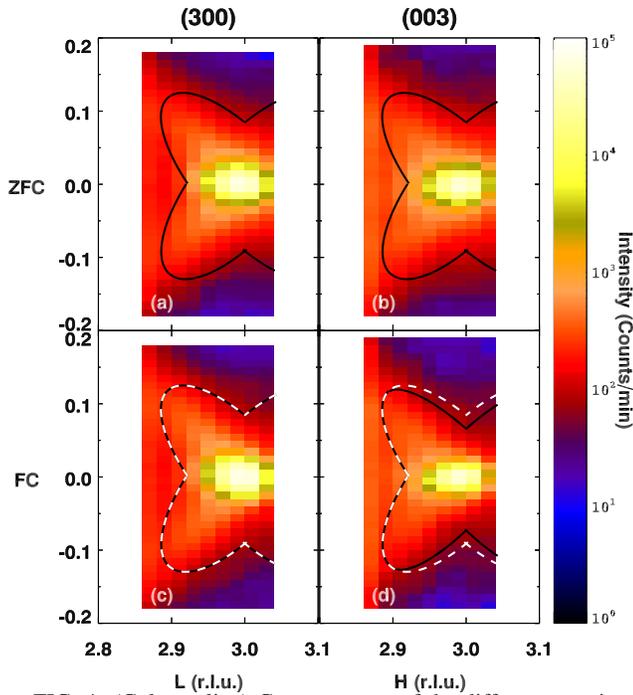}\qquad
\caption{(Color online) Contour maps of the diffuse scattering
measured near (300) and (003) at 400\,K under both ZFC and FC
conditions. The solid and dashed lines are guides to the eyes.}
\label{fig:3}
\end{figure}

\begin{figure}[ht]
\includegraphics[width=\linewidth]{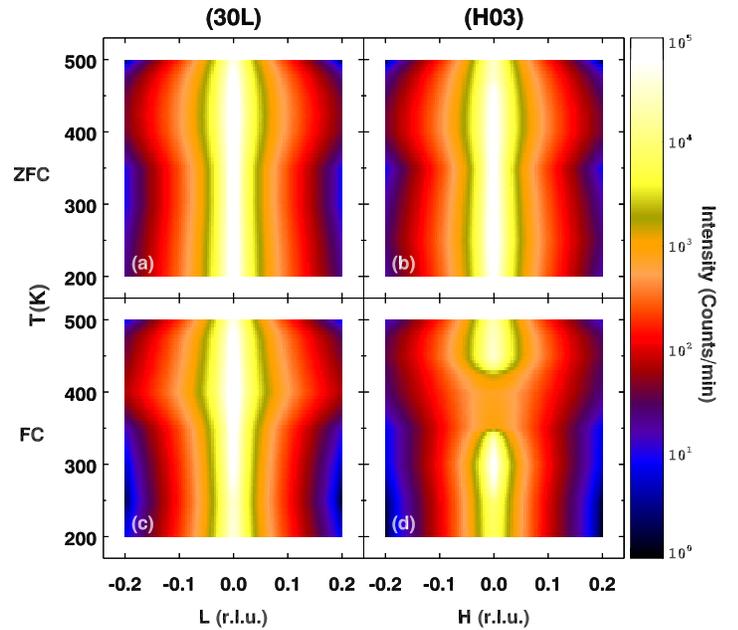}
\caption{(Color online) Contour maps of the fitted diffuse
scattering intensities versus temperature along directions
transverse to (300) and (003) under both ZFC and FC conditions .}
\label{fig:4}
\end{figure}

\begin{figure}[ht]
\includegraphics[width=\linewidth]{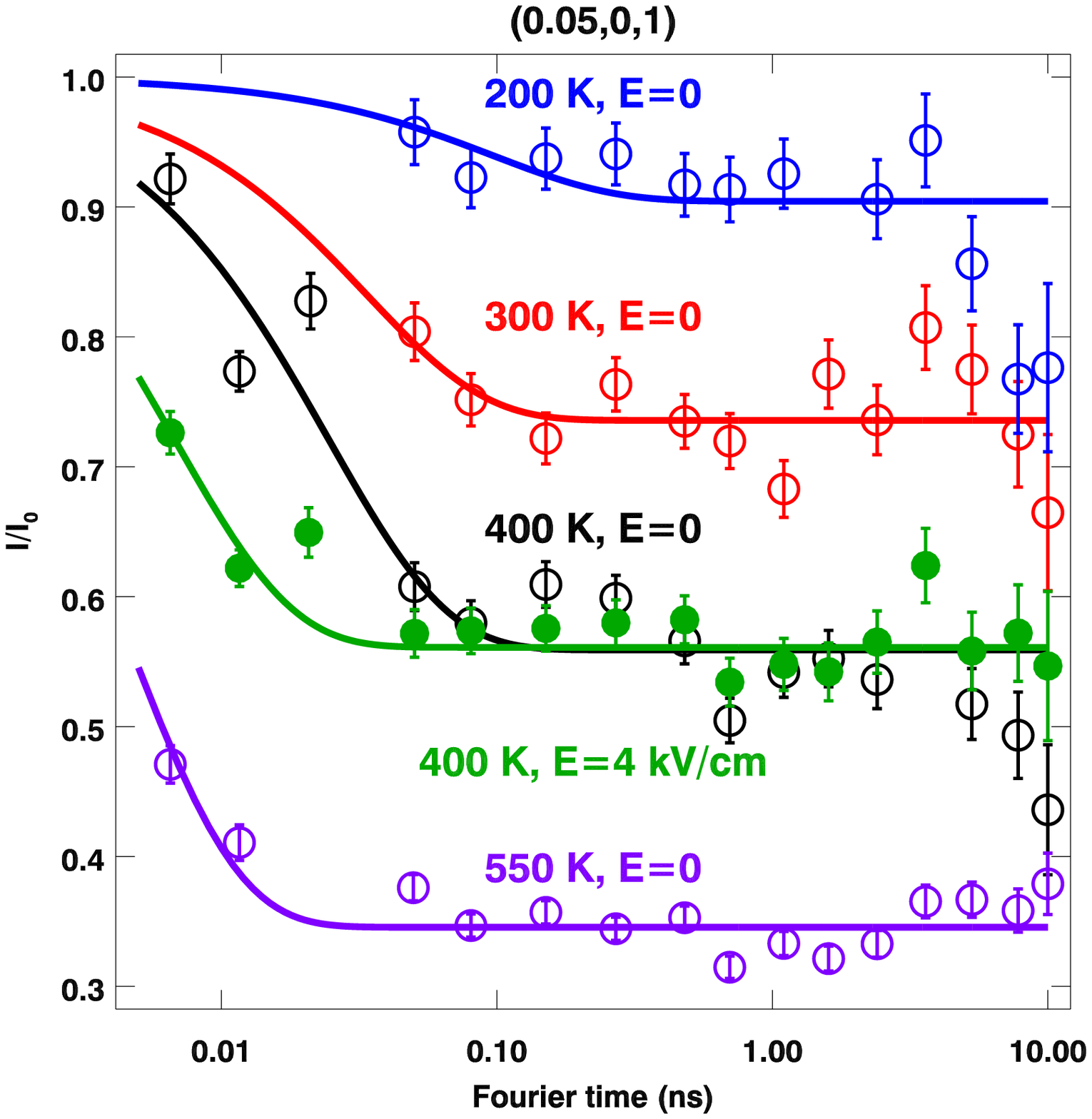}
\caption{(Color online) Neutron spin-echo data for different
temperatures and electric field effect at 200\,K (blue); 300\,K
(red); 400\,K (black; green); 550\,K (purple).  The error bars were
determined from the square root of I and I$_{0}$. All data were
taken under ZFC conditions shown with open circles except for the
400\,K FC data shown with closed symbols.  Lines are least square
fits to the data based on the model described in the text.}
\label{fig:5}
\end{figure}

\section{Experimental Details}

The sample used in our experiment is a PZN-4.5\%PT single crystal
purchased from TRS ceramics.  The crystal is rectangular with
\{100\} cut surfaces and dimensions of
10$\times$10$\times$3\,mm$^3$. The sample has a cubic lattice
spacing of $a = 4.05$\,\AA\ at 300\,K; thus 1\,rlu (reciprocal
lattice unit) equals $2\pi/a = 1.55$\,\AA$^{-1}$.  Cr/Au electrodes
were sputtered onto the two largest opposing crystal surfaces. The
Curie temperature of this compound $T_C \sim 475$\,K, which is
accompanied by a strong release of extinction that is visible at the
(100) Bragg peak and shown in panel (a) of Fig.~\ref{fig:2.5}.
Neutron diffuse scattering measurements were performed on the BT-9
triple-axis spectrometer located at the NIST Center for Neutron
Research (NCNR).  We used horizontal beam collimations of
40$'$-47$'$-sample-40$'$-80$'$ and a fixed final neutron energy of
14.7\,meV ($\lambda = 2.359$\AA). Two pyrolytic graphite filters
were placed before and after the sample to minimize the presence of
neutrons with higher-order wavelengths.  An external electric field
$E=4$\,kV/cm was applied along [001] above 550\,K during all of the
field-cooled (FC) measurements.

Neutron spin-echo measurements were performed on the NG-5 NSE
spectrometer, also located at the NCNR. The experiment was performed
at the scattering vector ${\bf Q}$=(0.05,0,1) ($|{\bf Q}|=1.559
$\,\AA$^{-1}$) and at a neutron wavelength $\lambda=5.5$\,\AA ~for
all time scales.  The instrumental resolution was measured with the
sample cooled to 40\,K where all dynamical processes occurring on
instrumentally accessible time scales are assumed to be frozen.

\section{Results and Discussion}

\subsection{Evidence of the T1-diffuse scattering component}\label{TAS data}

Our study of the T1-diffuse scattering was conducted by performing a
series of linear $q$-scans, represented by the dashed arrows drawn
in panel (c) of Fig.~\ref{fig:1}, under both field-cooled (FC) and
zero-field-cooled (ZFC) conditions around the (300), (003), (202),
and (101) Bragg peaks.  Diffuse scattering intensities measured at
(3,0,0.06) and (-0.06,0,3) are plotted in panel (b) of
Fig.~\ref{fig:2.5}. The temperature dependence similar to that
reported for the T2-diffuse scattering in that the diffuse
scattering increases with cooling and saturates at low temperature.
The peak observed near $T_C$ is most likely the result of the
critical scattering that appears near the structural phase
transition; similar behavior was observed by Stock {\it et al.} in
PZN.~\cite{Stock2004PRB}  Linear $q$-scans of the diffuse scattering
intensity measured at 400\,K under both FC and ZFC conditions are
plotted in Fig.~\ref{fig:2}. It is immediately clear that the
diffuse scattering intensities are suppressed around (003) after
field cooling whereas no such change occurs around (300); this
effect is also evident in panel (b) of Fig.~\ref{fig:2.5}. These
results are consistent with those obtained previously on
PZN-8PT.~\cite{Gehring2004prb}  At this point it is important to
recall that the neutron diffuse scattering cross section resulting
from correlated ionic displacements is proportional to $|{\bf
Q}\cdot{\bf \epsilon}|^2$, where $\bf \epsilon$ is a unit vector
along the displacement direction (for a more detailed discussion see
Section.~\ref{Calculation}). Therefore measurements of the diffuse
scattering intensity made near (300) are mainly sensitive to
short-range ordered ionic displacements oriented along [100] while
those made near (003) reflect the presence of ordered ionic
displacements oriented along [001].  The data shown in
Fig.~\ref{fig:2} therefore suggests that the short-range ordered
ionic displacements oriented along [001] are significantly
suppressed by an external electric field applied along [001].  This
scattering is what we refer to as T1-diffuse scattering.

Mesh scans were also performed to map out the geometry of the
diffuse scattering intensity distributions around various Bragg
peaks under different conditions.  The resulting intensity contours
measured near (300) and (003) at 400\,K are displayed in
Fig.~\ref{fig:3}. The white, dashed lines shown in panels
 (c) and (d) of Fig.~\ref{fig:3} are guides to the eye that describe the
shape of constant diffuse scattering intensity contours (which
include both T1 and T2 components) under ZFC conditions , as shown
in panels (a) and (b). The corresponding intensity contour maps
measured under FC conditions are shown in panels (c) and (d). Near
(300), the FC mesh scan reveals a shape that is very similar to that
obtained under ZFC conditions; however the one measured near (003)
appears to be slightly narrower, but only in the direction
transverse ([100]) to the scattering vector $\textbf{Q} = (003)$.
This asymmetry is important because it implies that the T1-diffuse
scattering associated with short-range ordered [001] ionic
displacements is distributed primarily along the transverse
direction ([100]) near (003) but not longitudinally (i.\~e.\ not
along [001]). This is why we call this scattering ``T1-diffuse
scattering''; both its polarization and distribution in $q$-space
resemble that of the T1 phonon mode.  On the other hand, the
butterfly-shaped T2-diffuse scattering, which is strongly affected
by an electric field applied along [111], is \emph{not} affected by
an electric field applied along [001].  Apparently the T2-diffuse
scattering dominates the diffuse scattering intensity measured at
most $q$ values; the presence of the T1-diffuse scattering component
only becomes evident through a change in its intensity once an
external electric field is applied along [001].

Our measurements near (300) and (003) suggest that an [001]-oriented
electric field can only reduce the diffuse scattering intensities
associated with ionic displacements that are parallel to the field.
To test this idea further, we studied the diffuse scattering
intensity near (101) as well.  Unlike the situation near (300) and
(003), both [100] and [001]-oriented atomic
displacements/polarizations will contribute to the overall neutron
diffuse scattering cross section near (101) because then $\textbf{Q}
\parallel [101]$.  Linear scans made along the $L$ ([001]) and $H$
([100]) directions near (101) are shown in panels (c) and (d) of
Figs.~\ref{fig:2}.  As shown in panel (d) of Fig.~\ref{fig:2}, the
field suppresses the diffuse scattering intensity distributed along
$\textbf{\emph{q}} \parallel [100]$, which must come from ionic
displacements oriented along [001], as was the case for the
T1-diffuse scattering measured near (003).  By symmetry there must
also be diffuse scattering associated with ionic displacements
oriented along [100], which is distributed along
\textbf{\emph{q}}$\parallel [001]$.  However, as was the case near
(300), this part of the T1-diffuse scattering is not affected by the
external [001] field as is shown in panel (c) of Fig.~\ref{fig:2}.

The temperature dependence of the field-induced suppression of the
T1-diffuse scattering is shown in Fig.~\ref{fig:4}. Transverse
$q$-scans across (300) and (003) were measured between 200\,K and
500\,K.  These scans were fit to a resolution-limited Gaussian
function of $q$, used to describe the Bragg peak intensity, and a
broad Lorentzian function of $q$, which describes the diffuse
scattering intensity.  In Fig.~\ref{fig:4} only the fitted diffuse
scattering intensities are plotted versus temperature and $q$ and
converted into color contour maps.  There are only tiny differences
between the ZFC (see panel (a)) and FC (see panel (c)) measurements
made near (300); however the diffuse scattering intensities near
(003) shown in panel (d) are strongly suppressed in the FC condition
for $T < T_C$ compared to those measured under ZFC conditions (panel
(b)).  This suppression seems to be largest for temperatures between
$T_C$ and $\sim400$\,K, and becomes less pronounced at lower
temperatures.  This can be understood if the PNR gradually freeze
with cooling and become harder to be influenced by an external
field.

\subsection{Static versus dynamic origin of the T1-diffuse scattering component}

Another question that has yet to be answered is whether PNR have a
static or dynamic origin.  Previous work using cold neutron
spectrometers, which provide significantly better energy resolution
than do thermal neutron spectrometers, have shown that the onset of
elastic (static) diffuse scattering occurs at much lower
temperatures than previously believed, i.~e.\, well below the Burns
temperature
$T_d$.~\cite{Hiraka2004prb,Rotaru2008,Gehring2009,Stock2010}  These
results imply that at high temperature the diffuse scattering is (at
least partially) dynamic in nature.  In order to better probe the
energy/time scale of the diffuse scattering, we have performed
spin-echo measurements on the same PZN-4.5\%PT single crystal at the
reciprocal lattice point (0.05,0,1.0) at different temperatures. The
diffuse scattering intensities shown in Fig.~\ref{fig:5} have been
corrected for the instrumental resolution and plotted in the form of
$I(\textbf{Q},t)/I_0(\textbf{Q},0)$ versus Fourier time $t$. These
intensity plots can in fact provide information on how much of the
total diffuse scattering at this \textbf{Q} is static. One sees that
at high temperature ($T=550$\,K), the diffuse scattering intensity
decays to $I/I_0 \alt 40\%$ within $\alt 0.01$\,ns. In other words,
less than 40\% of the diffuse scattering is static, while the rest
of the intensity is dynamic in nature having a relaxation time less
than $0.01$\,ns (see Eqn.~\ref{eqn:4}), which corresponds to an
energy half width of $\hbar\Gamma\gtrsim 0.066$\,meV. On cooling the
relative size of the static diffuse scattering component increases,
which is consistent with freezing of the PNR. At 200\,K, the diffuse
scattering is almost completely static ($I/I_0 \agt 90\%$). These
results are qualitatively similar to those obtained on pure PMN by
Stock {\it et al}.~\cite{Stock2010}

We have also studied the effect of an external field along [001] on
the diffuse scattering measured at 400\,K.  At (0.05,0,1) we should
be sensitive to the T1-diffuse scattering associated with [001]
ionic displacements, which should be partially suppressed by the
[001] field.  We find that, although the overall diffuse scattering
intensity is reduced by the field, the static ratio $I/I_0$ at large
Fourier times is not affected.  However, the dynamic component
decays much faster when the sample is FC compared to ZFC.  The lines
through the data shown in Fig.~\ref{fig:5} are based on fits to a
one parameter decay function
\begin{equation}\label{eqn:4}
I(t)/I(0)=S+(1-S)\exp(-t/\tau)
\end{equation}
In Table~\ref{tab:3}, we list the temperature dependence of these
parameters, including the static fraction $S$, and the relaxation
time $\tau$, as well as the energy half width at half maximum
$\hbar\Gamma=1/\tau$. At 400\,K, $\tau \approx 0.025$\,ns under ZFC
conditions and $\tau \approx 0.0068$\,ns under FC conditions.  These
$\tau$ values correspond to energy widths of $\hbar\Gamma\approx
0.027$\,meV (ZFC) and $\hbar\Gamma\approx 0.10$\,meV (FC),
respectively.

\begin{table}
\caption{ Values of the fitting parameters used in Eqn.~\ref{eqn:4}.
The energy width (half width at half maximum, HWHM) is calculated
from $E =\hbar\Gamma=\hbar/\tau$ }
\begin{ruledtabular}
\begin{tabular}{lcccccc}
  \quad & 550 K & FC 400 K & ZFC 400 K & 300 K & 200 K   \\
  \hline
Static fraction   (S/$\%$) & 35 & 56 & 56 & 74 & 90  \\
Relaxation time   ($\tau$/ns) & 0.0042 & 0.0068 & 0.025 & 0.033 & 0.097  \\
Energy width      (E/meV) & 0.16 & 0.1 & 0.027 & 0.02 & 0.007  \\
\end{tabular}
\end{ruledtabular}
\label{tab:3}
\end{table}

Although we have obtained the relative fraction of the static
portion of the diffuse scattering intensity in PZN-4.5\%PT, it is
clear that these spin-echo measurements do not provide the best
time/energy scale to probe the relaxation time of the dynamic component of
the diffuse scattering. To get better results one would need to
perform measurements with energy resolution of about 20\,$\mu$eV.
Future measurements using the neutron backscattering technique are
being planned for this purpose.

\subsection{Short-range correlated ionic displacements associated with the T1-diffuse scattering}
\label{Calculation}

In general, the diffuse scattering intensity $I_{diff}(\textbf{Q})$
coming from a single source at $\textbf{Q}=\textbf{G}+\textbf{q}$
near the Bragg peak $\textbf{G}$ can be approximately described by
~\cite{gxu2006prb}
\begin{equation}\label{eqn:1}
    I_{diff}(\textbf{Q})=A |F_{diff}(\textbf{G})|^2 |f(\textbf{q})|^2.
\end{equation}
Here $A$ is a scale factor. $|f(\textbf{q})|^2$ is the Fourier
transform of the real space shape of the PNR and describes the shape
of the diffuse scattering intensity distribution around a Bragg
peak.  This term is in principle independent of Brillouin zone.
$|F_{diff}(\textbf{G})|^2$ is the diffuse scattering structure
factor and is Brillouin zone dependent; it can be written as:
\begin{equation}\label{eqn:2}
    |F_{diff}(\textbf{G})|^2 =
|\sum_i \textbf{Q}\cdot{\bf \xi_{i}} b_i \exp{(-W_i)}
\exp{(i\textbf{G}\cdot \textbf{R}_i})|^2 ,
\end{equation}
where ${\bf \xi_{i}}, b_i$, and $R_i$ are the ionic displacement
vector, neutron scattering length, and the lattice position of the
$i$th atom in the unit cell, respectively, and $\exp(-W_i)$ is the
Debye-Waller factor.

Many sources contribute to the total neutron diffuse scattering
cross section in relaxors. In the case of PZN-4.5\%PT these include
six $\langle110\rangle$ polarization directions and three
$\langle001\rangle$ polarization directions associated with the T2
and T1-diffuse scattering cross sections, respectively.  It is very
difficult to determine the absolute intensity of the T1-diffuse
scattering directly because it coexists with the much stronger
T2-diffuse scattering.  However, we have shown that a [001]-oriented
electric field can suppress portion of the T1-diffuse scattering
intensity that is specifically associated with [001]-oriented ionic
displacements while leaving that associated with [100] and [010]
ionic displacements (and the T2-diffuse scattering) unaffected. One
can then exploit this fact and measure the change in the T1-diffuse
scattering cross section in different Brillouin zones to determine
the corresponding structure factor and solve for the relative
magnitudes of the ionic shifts that contribute to the T1-diffuse
scattering.  If one assumes that the Debye-Waller factor does not
vary appreciably with Brillouin zone, then one can simplify
Eqn.~\ref{eqn:2} to~\cite{gxu2004prb}
\begin{equation}\label{eqn:3}
    |F_{diff}(\textbf{G})|^2 \propto
|\textbf{Q}\cdot {\bf \epsilon}|^2|\sum_i b_i\cdot\xi_i
\exp{(i\textbf{G}\cdot \textbf{R}_i)}|^2
\end{equation}
where, again, $\bf \epsilon$ is the unit vector along the ionic
displacement (polarization) direction.

As discussed in section~\ref{TAS data}, the T1-diffuse scattering
associated with short-range ordered, [001]-oriented, ionic
displacements is distributed along [100] across each Bragg peak in
the (H0L) plane.  We therefore chose to measure T1-diffuse
scattering intensities at the reciprocal lattice positions
$(q,0,3)$, $(1+q,0,1)$, and $(2+q,0,2)$.  The differences between
ZFC and FC measurements made at $q=0.06$\,rlu, shown in
Table.~\ref{tab:1}, are then used as the relative structure factors
for the T1-diffuse scattering cross section.  Here we have assumed
that the electric field does not affect the structure factors.  This
is equivalent to the assumption that the electric field reduces the
T1-diffuse scattering uniformly independent of Brillouin zone.
Knowing $|F_{diff}(\textbf{G})|^2$ at different Bragg peaks, one can
solve for the average Pb, Zn/Nb/Ti, and O displacements in the unit
cell that contribute to the T1-diffuse scattering. Similar to what
was done in Ref.~\onlinecite{gxu2006prb}, we neglect any possible
distortion or rotation of the oxygen octahedra and assume that all
six oxygens in one unit cell move as a unit. The relative ionic
shifts that contribute to the T1-diffuse scattering intensities are
listed in Table.~\ref{tab:2}. In Table.~\ref{tab:2} we have
decomposed these ionic shifts into different components, i.\ e.\ one
acoustic/strain component that corresponds to the uniform phase
shift in which all atoms in the unit cell move
together,~\cite{hirota2002} and two optic components that correspond
to the Slater and Last modes~\cite{Shirane1970,gxu2006prb} where the
atoms in the unit cell only move relative to each other with no
change in the center of mass. Apparently, as was the case for the
T2-diffuse scattering, the local ionic displacements that give rise
to the T1-diffuse scattering also consist of both acoustic/strain
and optic/polar components.

We have also examined the T2-diffuse scattering intensity by making
measurements at reduced wavevectors $q \parallel [110]$ offset from
various Bragg peaks.  The structure factors and ionic displacements
contributing to the T2-diffuse scattering obtained from our
PZN-4.5\%PT sample are very similar to those obtained previously for
PZN-8\%PT (see Tables I and II in Ref.~\onlinecite{gxu2006prb}). We
can therefore compare the local ionic structures that give rise to
the T1-diffuse scattering to those that produce the T2-diffuse
scattering. We find that the ionic displacements associated with the
T1-diffuse scattering have a larger optic component;  note that the
shifts for the Zn/Nb/Ti site associated with the T1-diffuse
scattering are larger than those associated with the T2-diffuse
scattering in Ref.~\onlinecite{gxu2006prb}. Previous work has shown
that the T2-diffuse scattering is coupled strongly to transverse
acoustic phonons and strains in relaxor
systems,~\cite{gxu2008nm,Stock2005} but not with the soft transverse
optic (TO) phonon.  We expect that this situation could be different
for the T1-diffuse scattering, which may in fact couple more
strongly to the soft TO phonon because it has a larger optic
component. This will of course need to be verified by future
experiments.

\begin{table}

\caption{Values for $|\textbf{Q}\cdot {\bf \epsilon}|^2$ and the
difference (ZFC$-$FC) of the T1-diffuse scattering intensity
measured at $(3+q,0,0)$, $(2+q,0,2)$, and $(1+q,0,1)$ for
$q=0.06$\,r.l.u.}
\begin{ruledtabular}
\begin{tabular}{cccc}
  \quad & (300) & (202) & (101) \\
  \hline
  $|\textbf{Q}\cdot {\bf \epsilon}|^2$ & 9 & 8 & 2 \\
  500 K & 13 & 4 & 9 \\
  400 K & 320 & 72  & 115 \\
  300 K & 154 & 1 & 67 \\
  200 K & 193 & 0 & 75 \\
\end{tabular}
\end{ruledtabular}

\label{tab:1}

\end{table}

\begin{table}
\caption{Calculated T1-type ionic displacements and relative
magnitudes of different modes.  All displacements have been
normalized to that for the Pb cation. The values for the Slater,
Last and Shift modes are based on the oxygen octahedra
displacements.}
\begin{ruledtabular}
\begin{tabular}{ccccccc}
  \quad & $\delta_{Pb}$ & $\delta_{Zn,Nb}$ & $\delta_{O}$ & Shift & Slater & Last \\
  \hline
  500 K & 1.0 & 0.54 & -0.44 & 0.68 & -0.60 & -0.52 \\
  400 K & 1.0 & 0.22 & -0.33 & 0.63 & -0.34 & -0.62 \\
  300 K & 1.0 & 0.35 & -0.61 & 0.62 & -0.59 & -0.63 \\
  200 K & 1.0 & 0.30 & -0.65 & 0.60 & -0.58 & -0.65 \\
\end{tabular}
\end{ruledtabular}
\label{tab:2}
\end{table}

\section{Summary}

Our neutron scattering measurements clearly show that in addition to
the well-known butterfly-shaped diffuse scattering, a second,
distinct, diffuse scattering component also exists in the PZN-$x$PT
relaxor system.  This so-called T1-diffuse scattering can be
differentiated from the butterfly-shaped T2-diffuse scattering
through its dependence on an $\langle001\rangle$-oriented electric
field.  Quantitative analysis of the short-range ordered,
$\langle001\rangle$-oriented, ionic displacements associated with
the T1-diffuse scattering suggest that both acoustic/strain and
optic/polar ionic displacements are present, a situation that is
very similar to that for the T2-diffuse scattering. Spin-echo
measurements also show that the diffuse scattering cross section
exhibits a large dynamic component at high temperature paraelectric
phase above $T_C$ and gradually freezes, becoming almost entirely
static for temperatures well below $T_C$ in the ferroelectric phase.

The electric field dependence of the T1 and T2-diffuse scattering
cross sections are quite different; this implies that they might
originate from independent nano-scale polar structures. On the other
hand, they are very similar in many other aspects:  both exhibit
strain/polar components and both freeze with cooling.  This raises
another scenario in which the two diffuse scattering components
might be associated within different $\langle001\rangle$ and
$\langle110\rangle$ components of the ionic displacements with the
same nano-scale polar structure. The average local atomic shifts in
these nano-scale structures could very well be along other
directions, e.g., along $\langle111\rangle$ directions as suggested
by neutron Pair Distribution Function (PDF) measurements from
PMN~\cite{Jeong2005}. The neutron diffuse scattering intensities
that we observe here arise only from one or more components of these
local displacements that become spatially short-range ordered. Any
components of atomic displacements that are entirely disordered will
only contribute to the overall background and will not affect the
diffuse scattering intensities discussed in this work. Our data do
not allow us to determine definitively which scenario is correct,
thus more detailed studies are clearly required.  However, our
results do strongly suggest that when studying PNR in relaxor
systems, the existence of complex nano-scale polar structures
composed of both $\langle001\rangle$ and
$\langle110\rangle$-oriented ionic displacements will have to be
carefully taken into consideration; indeed, these may affect the
lifetimes of phonons propagating along these two sets of directions.

\section{Acknowledgments}

We wish to thank W.~Ratcliff, S.~M.~Shapiro, and S.~B.~Vakhrushev
for useful discussions.  Financial support from the US Department of
Energy under contract No.\ DE-AC02-98CH10886 and the Natural Science
and Engineering Research Council of Canada (NSERC) is also
gratefully acknowledged. The identification of any commercial
product or trade name does not imply endorsement or recommendation
by the National Institute of Standards and Technology.


\end{document}